\documentclass[twocolumn]{aastex631}
\usepackage{natbib}
\let\tablenum\relax
\usepackage{siunitx}
\usepackage{amsmath}
\usepackage[caption = false]{subfig}
\usepackage{xcolor}

\DeclareSIUnit\parsec{pc}

\graphicspath{{./}{figures/}}

\begin{document}

\title{The Local Galactic Transient Survey Applied to an Optical Search for Directed Intelligence}

\email{agthomas@ucsb.edu}
\author[0000-0001-9528-8147]{Alex Thomas}
\affiliation{University of California Santa Barbara \\
Santa Barbara, CA 93106, USA}
\author[0000-0002-2249-0595]{Natalie LeBaron}
\affiliation{University of California Santa Barbara \\
Santa Barbara, CA 93106, USA}
\author{Luca Angeleri}
\affiliation{University of California Santa Barbara \\
Santa Barbara, CA 93106, USA}
\author{Phillip Morgan}
\affiliation{University of Colorado Boulder \\
Boulder, CO 80309, USA}
\author{Varun Iyer}
\affiliation{University of California Santa Barbara \\
Santa Barbara, CA 93106, USA}
\author[0009-0008-0190-3179]{Prerana Kottapalli}
\affiliation{University of California Santa Barbara \\
Santa Barbara, CA 93106, USA}
\author{Enda Mao}
\affiliation{University of California Santa Barbara \\
Santa Barbara, CA 93106, USA}
\author[0000-0002-6836-181X]{Samuel Whitebook}
\affiliation{University of California Santa Barbara \\
Santa Barbara, CA 93106, USA}
\author{Jasper Webb}
\affiliation{University of California Santa Barbara \\
Santa Barbara, CA 93106, USA}
\author{Dharv Patel}
\affiliation{University of California Santa Barbara \\
Santa Barbara, CA 93106, USA}
\author{Rachel Darlinger}
\affiliation{University of California Santa Barbara \\
Santa Barbara, CA 93106, USA}
\author{Kyle Lam}
\affiliation{University of California Santa Barbara \\
Santa Barbara, CA 93106, USA}
\author{Kelvin Yip}
\affiliation{University of California Santa Barbara \\
Santa Barbara, CA 93106, USA}
\author{Michael McDonald}
\affiliation{University of California Santa Barbara \\
Santa Barbara, CA 93106, USA}
\author{Robby Odum}
\affiliation{University of California Santa Barbara \\
Santa Barbara, CA 93106, USA}
\author{Cole Slenkovich}
\affiliation{University of California Santa Barbara \\
Santa Barbara, CA 93106, USA}
\author{Yael Brynjegard-Bialik}
\affiliation{University of California Santa Barbara \\
Santa Barbara, CA 93106, USA}
\author{Nicole Efstathiu}
\affiliation{University of California Santa Barbara \\
Santa Barbara, CA 93106, USA}
\author{Joshua Perkins}
\affiliation{University of California Santa Barbara \\
Santa Barbara, CA 93106, USA}
\author{Ryan Kuo}
\affiliation{University of California Santa Barbara \\
Santa Barbara, CA 93106, USA}
\author{Audrey O'Malley}
\affiliation{University of California Santa Barbara \\
Santa Barbara, CA 93106, USA}
\author{Alec Wang}
\affiliation{University of California Santa Barbara \\
Santa Barbara, CA 93106, USA}
\author{Ben Fogiel}
\affiliation{University of California Santa Barbara \\
Santa Barbara, CA 93106, USA}
\author{Sam Salters}
\affiliation{University of California Santa Barbara \\
Santa Barbara, CA 93106, USA}
\author{Marlon Munoz}
\affiliation{University of California Santa Barbara \\
Santa Barbara, CA 93106, USA}
\author{Natalie Kim}
\affiliation{University of California Santa Barbara \\
Santa Barbara, CA 93106, USA}
\author{Lee Fowler}
\affiliation{University of California Santa Barbara \\
Santa Barbara, CA 93106, USA}
\author{Ruiyang Wang}
\affiliation{University of California Santa Barbara \\
Santa Barbara, CA 93106, USA}
\author{Philip Lubin}
\affiliation{University of California Santa Barbara \\
Santa Barbara, CA 93106, USA}

\begin{abstract}
\noindent We discuss our transient search for directed energy systems in local galaxies, with calculations indicating the ability of modest searches to detect optical Search for Extraterrestrial Intelligence (SETI) sources in the closest galaxies. Our analysis follows \cite{Lubin2016} where a messenger civilization follows a beacon strategy we call ``intelligent targeting.'' We plot the required laser time to achieve an SNR of 10 and find the time for a blind transmission to target all stars in the Milky Way to be achievable for local galactic civilizations. As high cadence and sky coverage is the pathway to enable such a detection, we operate the Local Galactic Transient Survey (LGTS) targeting M31 (the Andromeda Galaxy), the Large Magellanic Cloud (LMC), and the Small Magellanic Cloud (SMC) via Las Cumbres Observatory's (LCO) network of 0.4\,m telescopes. We explore the ability of modest searches like the LGTS to detect directed pulses in optical and near-infrared wavelengths from Extraterrestrial Intelligence (ETI) at these distances and conclude a civilization utilizing less powerful laser technology than we can construct in this century is readily detectable with the LGTS's observational capabilities. Data processing of 30,000 LGTS images spanning 5 years is in progress with the TRansient Image Processing Pipeline (TRIPP; \cite{TRIPP}).
\end{abstract}

\keywords{SETI, Search for Extra Terrestrial Intelligence, Laser Optics, Phased Arrays}

\section{Introduction} \label{Introduction}
\noindent The Search for Extraterrestrial Intelligence (SETI) seeks to discover extraterrestrial communications based on our understanding of terrestrial technologies and potential future capabilities. Originally, this search focused on radio frequencies, which were well understood as means of communication when SETI efforts began in the 1960s (\cite{Tarter2001}). However, within a year of the discovery of lasers, \cite{Schwartz1961} proposed beam-like signals in optical wavelengths as a means of interstellar communication. Indeed, the continually growing power output of laser technology in recent decades suggests that lasers may be an effective means of interstellar communication (\cite{Howard2004}; \cite{Townes1983}). Thus, examining optical and near-infrared frequencies for laser communications, which few SETI searches have examined (\cite{Hippke2018} and \cite{Price2020}) is a natural extension of SETI efforts.

Optical SETI (OSETI) initially focused on systems designed to observe individual stars and detect transient phenomena, such as brief, nanosecond-scale laser pulses, which could indicate technosignals (\cite{beskin1995}; \cite{Wright2001}; \cite{Howard2004}; \cite{Maire2016}). However, as the likelihood of any given star hosting detectable ETI is small, and since technosignals may not originate from around stars, wide sky coverage is essential to boost the probability of a detection.

The first wide-field OSETI survey was the Harvard/Planetary Society all-sky search which used a 1.8\,m optical telescope imaging a 1.6{\textdegree} × 0.2{\textdegree} field of view at Oak Ridge Observatory (\cite{Howard2007}). Since then, several next-generation instruments have been deployed. The Pulsed All-sky Near-infrared Optical SETI (PANOSETI; \cite{Wright2018}; \cite{Maire2022}) observatory will ultimately consist of 24, 0.46\,m telescopes (two of which are currently assembled at Lick Observatory) and aims to provide 2,350 $\unit{deg^2}$ of instantaneous sky coverage (7,000 times that of the Harvard/Planetary Society all-sky search). Additionally, the SETI Institute's LaserSETI\footnote{https://laserseti.net/instrument/}\textsuperscript{,}\footnote{https://www.seti.org/search-et-2022} program has deployed three of 12 planned wide-field slitless spectroscopes each surveying 4,395 $\unit{deg^2}$.

Exploring the possibility of civilizations attempting to communicate by using directed energy signals, \cite{Lubin2016} attempts to quantitatively and qualitatively describe what these signals might look like as a factor of the distance of the civilization and their level of technology. We follow an ``intelligent targeting'' assumption (\cite{Lubin2016}) which assumes messenger civilizations target the habitable zone of each stellar system with directed laser pulses rather than uniformly spreading transmission time within a target galaxy. This assumption boosts the probability of detection by orders of magnitude. While the intelligent targeting assumption has blind transmission and reception (we do not need to know the location of the messenger civilization and vice versa), the messenger civilization must possess a detailed knowledge of our galaxy's stellar motions and gravitational lensing at small angles in order to beam targeted systems. Assuming these prerequisites are met, we take the probability of a messenger civilization targeting a desired system with a directed laser to be near unity.

Due to the relatively small size of habitable zones for a given star (compared to the cross-sectional area between stars), this assumption boosts the probability of detection significantly. Taking the habitable zone of each system to be the diameter of earth's orbit, \cite{Lubin2016} finds a ${\sim}10^8$ increase in the probability of detection with a very conservative diameter of 10 AU (${\sim}$Saturn) yielding a ${\sim}10^6$ increase in the probability of detection. Following the intelligent targeting assumption, such a technosignal would appear to be a transient foreground point source and likely be non-periodic (in human timescales) due to sequentially targeting every Milky Way solar system.

In Section \ref{SETI} we discuss the expected flux, dwell time, SNR, and minimum laser time of a directed energy messenger civilization as calculated in \cite{Lubin2016}. We also discuss the ideal exposure time and its implications for SETI and METI (Messaging to extraterrestrial intelligence). In Section \ref{LGTS} we discuss our Local Galactic Transient Survey (LGTS) which utilizes 0.4 m telescopes from Las Cumbres Observatory Global Telescope network (LCOGT) to look for optical transient signals with short ({$\sim$}10 seconds) integration times. In Section \ref{conclusion} we give an overview of the status of LGTS and emphasize that LGTS+TRIPP could feasibly detect OSETI signals in LGTS data.

\section{Searching for Directed Energy} \label{SETI}
\noindent The power $P$(W) for a given civilization class $S$ as defined in \cite{Lubin2016} is
\begin{align}
P = F_e\epsilon_c10^{2S}. \label{power}
\end{align}
The civilization class $S$ is defined such that $10^S$ meters is the side length of a square laser array converting stellar power to laser power.
By this definition, a class 5 civilization is similar to a Kardashev Type I, while a class 11 civilization is similar to a Kardashev Type II (\cite{Kardashev}).
Our calculations make the following assumptions: solar illumination $F_e = \SI{1400}{W/\m\squared}$ (based on the solar illumination at the top of the Earth's atmosphere), and conversion efficiency of stellar power to laser power $\epsilon_c = 0.5$. 

Eq. \ref{power} neglects to include the small, wavelength-dependent attenuations from the interstellar mediums (ISMs) of the Milky Way and the messenger civilization, and the intergalactic medium (IGM). For example, the attenuation from the Milky Way towards M31 is $\sim0.17$ mag (\cite{Schlafly2011}; \cite{Dong2014}). For a messenger civilization located in the M31 bulge, the ISM attenuation from M31 is still small, $\lesssim0.5$ mag (\cite{Dong2014}). Similarly, the attenuation from the IGM is negligible for LGTS (\cite{Inoue2014}).

Using the power, we calculate the apparent and photon flux of a messenger civilization. The apparent flux $F(\unit{W\per\m\squared})$ of a laser emitted from a (luminosity) distance $L$(m) for a civilization of class $S$ and wavelength-dependent beam divergence solid angle $\Omega(\unit{sr}) = 4\lambda^210^{-2S} \unit{sr}$ is
\begin{align}
    F = \frac{P}{L^2\Omega} = \frac{F_e \epsilon_c 10^{4S}}{4 L^2 \lambda^2}.
\end{align}

\begin{figure}[hbt!]
    \centering
    \includegraphics[width = \columnwidth]{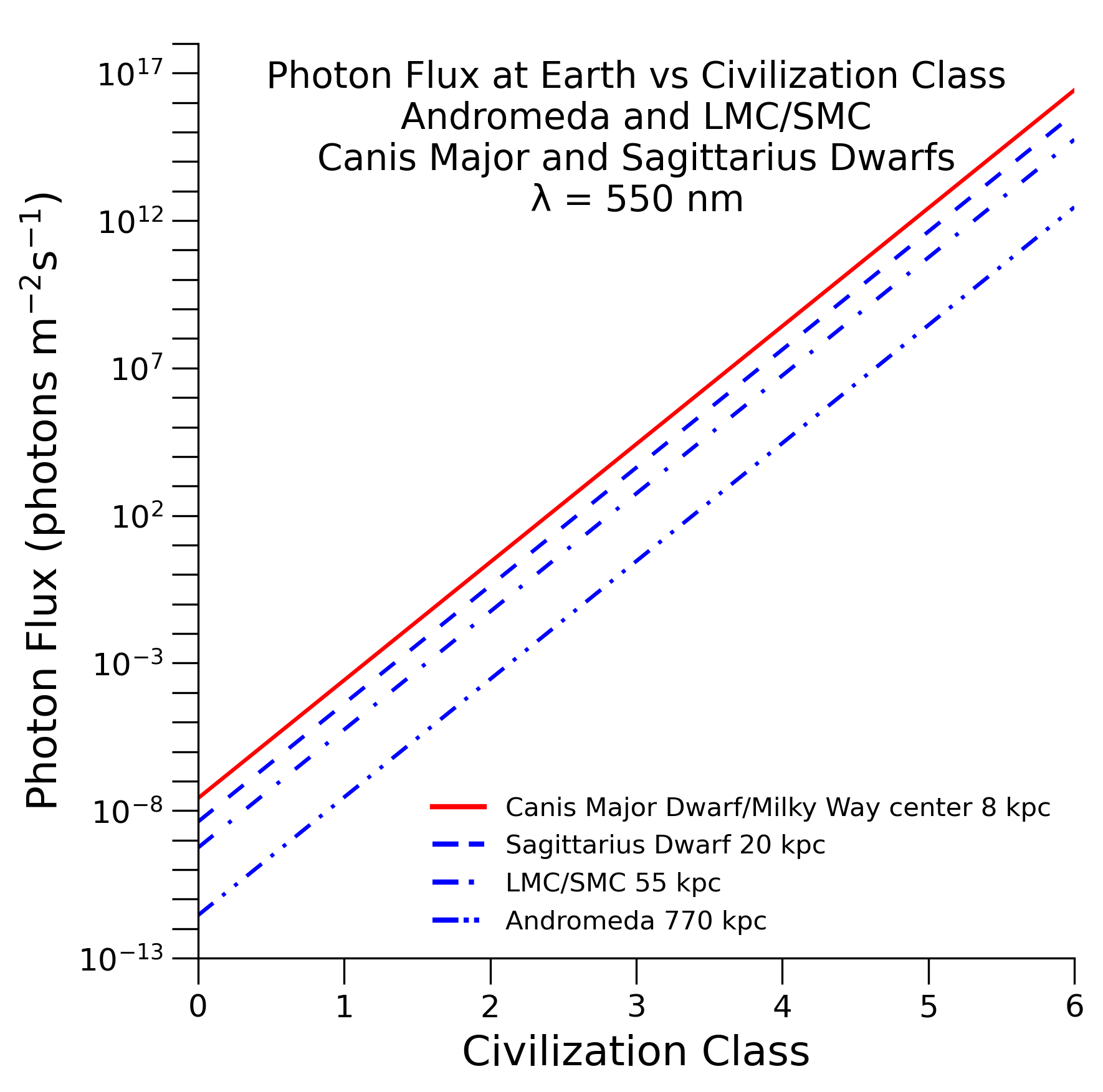}
    \caption{Expected photon flux incident at Earth emitted by various civilization classes and local galactic distances with wavelength $\SI{1.06}{\micro\m}$.}
    \label{Flux v Class}
\end{figure}

The apparent flux can be converted to photon flux in photons $\unit{\m^{-2}\s^{-1}}$ by dividing by the energy per photon in Joules (\cite{Lubin2016}). Figure \ref{Flux v Class} summarizes the photon flux versus civilization class for intragalactic sources and some nearby intergalactic galaxies.

{}

Exploring a class 4 civilization transmitting from M31, we take $S = 4$, $\lambda$ to be between $400$ nm and $700$ nm (optical wavelengths), and $L \approx 2.56 \pm 0.11$ Mly. These values give us approximate bounds, placing the photon flux between $\SI{2e4}{\gamma\m^{-2}\s^{-1}}$ and $\SI{4e4}{\gamma\m^{-2}\s^{-1}}$. From these flux values, the prospective apparent magnitude under these conditions is 16, without accounting for the ISM and IGM attenuation effects discussed above.

The spot size $s$ of the beam at the Earth is the product of the distance to the transmitter $L$ and the beam divergence full angle $\theta$,

\begin{equation}
s = L\theta = 2 L \lambda 10^{-S}.
\end{equation}

Using the prior constraints we find the approximate spot size to be ${\sim}10$ AU emphasizing the need for precise pointing, detailed knowledge of our galaxy's stellar motion, and gravitational lensing at small angles.

The dwell time $\tau_\textrm{dwell}$—how long a blindly transmitted beam would be visible on Earth—is the spot time divided by the relative transverse speed of the transmitter $V_t$
\begin{align}
\tau_\textrm{dwell} = \frac{s}{V_t} = \frac{2L\lambda} {10^SV_t}.
\end{align}
Using a typical transverse speed relative to Earth of 100 km/s to 1000 km/s, the dwell time bounds are $2 \times 10^6$ s to $4 \times 10^7$ s (\cite{Lubin2016}).
Given that the dwell time for an OSETI source is many orders of magnitude greater than our integration times, it is quite feasible for 0.4 m telescopes to capture any such signals within the dwell time.

\begin{figure}[h!]
    \centering
    \includegraphics[width = \columnwidth]{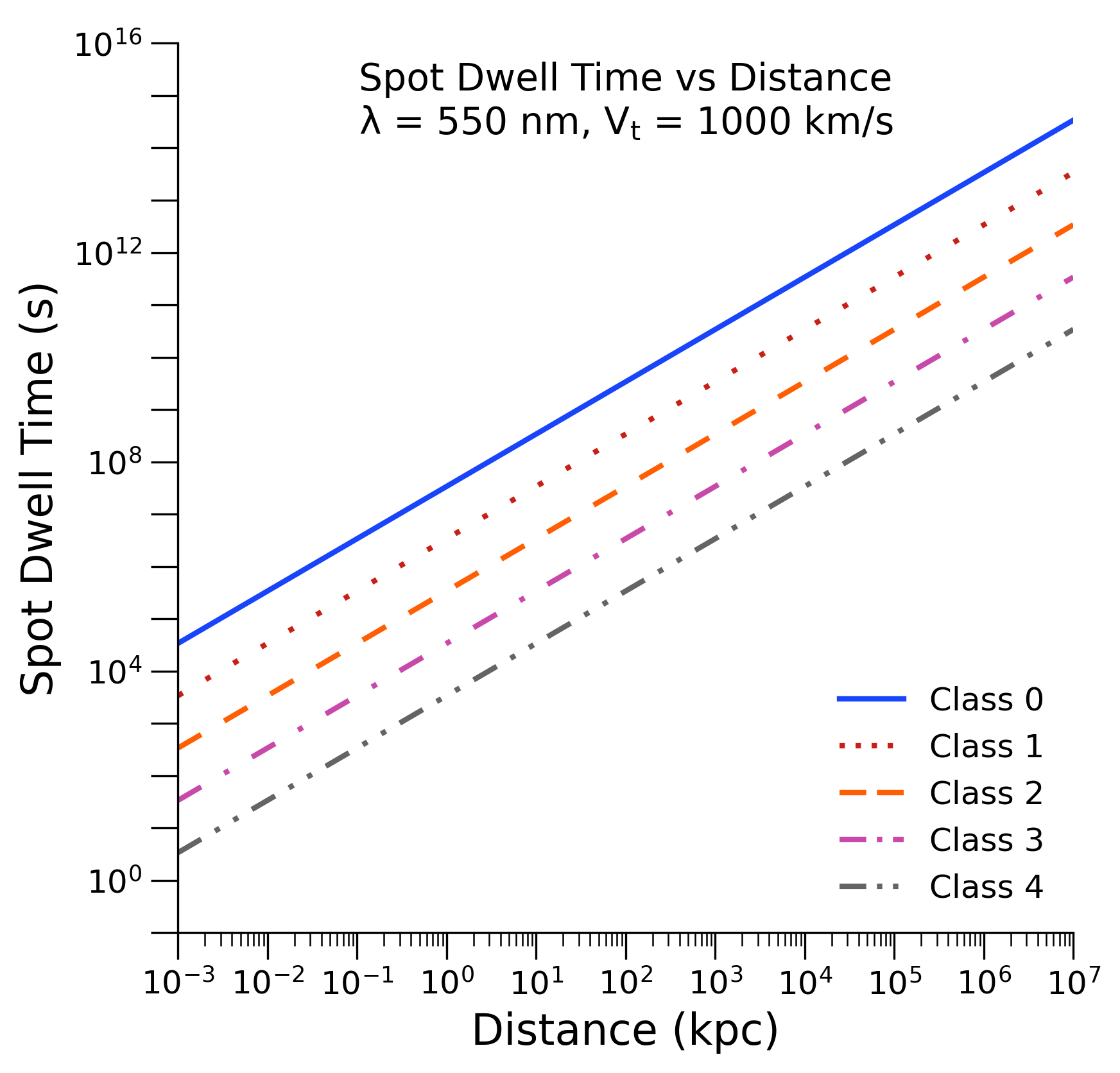}
    \caption{Spot dwell time vs distance. Lines are plotted for various civilization classes. Figure adapted with permission from \cite{Lubin2016}.}
    \label{Dwell Time}
\end{figure}

In Figure \ref{Dwell Time}, we report dwell time versus distance for various civilization classes
by assuming a Euclidean geometry for simplicity. 
As dwell time decreases with shorter distances and wavelengths, and higher transverse velocities, dwell time should only become an issue ($\tau_\textrm{dwell} < \tau_\textrm{SNR}$) for a high-class messenger civilization in the Milky Way. 

{}

Following \cite{Lubin2016}, the Signal-to-Noise Ratio (SNR) of a source computed relative to nearby pixels is

\begin{align}
    S_N \equiv \frac{S}{N} = \frac{FA_\epsilon \tau}{N_T} = \frac{FA_\epsilon \tau}{[N_R^2+ \tau(i_{DC} + F_BA_\epsilon \Omega)]^{1/2}}
\end{align}
where $F(\unit{\gamma\m^{-2}\s^{-1}})$ is photon flux, $A_\epsilon(\unit{\m\squared e^-\gamma^{-1}})$ is effective telescope area accounting for the quantum and optical efficiencies, $\tau(s)$ is integration time, $N_T(e^-)$ is the total noise, $N_R(e^-)$ is the readout noise, $i_{DC}(\unit{e^- \s^{-1}})$ is dark current, $F_B(\unit{\gamma\m^{-2}\s^{-1}sr^{-1}})$ is background flux per solid angle, and $\Omega(\text{sr})$ is beam divergence solid angle.

\begin{figure}[hbt!]
    \centering
    \includegraphics[width = \columnwidth]{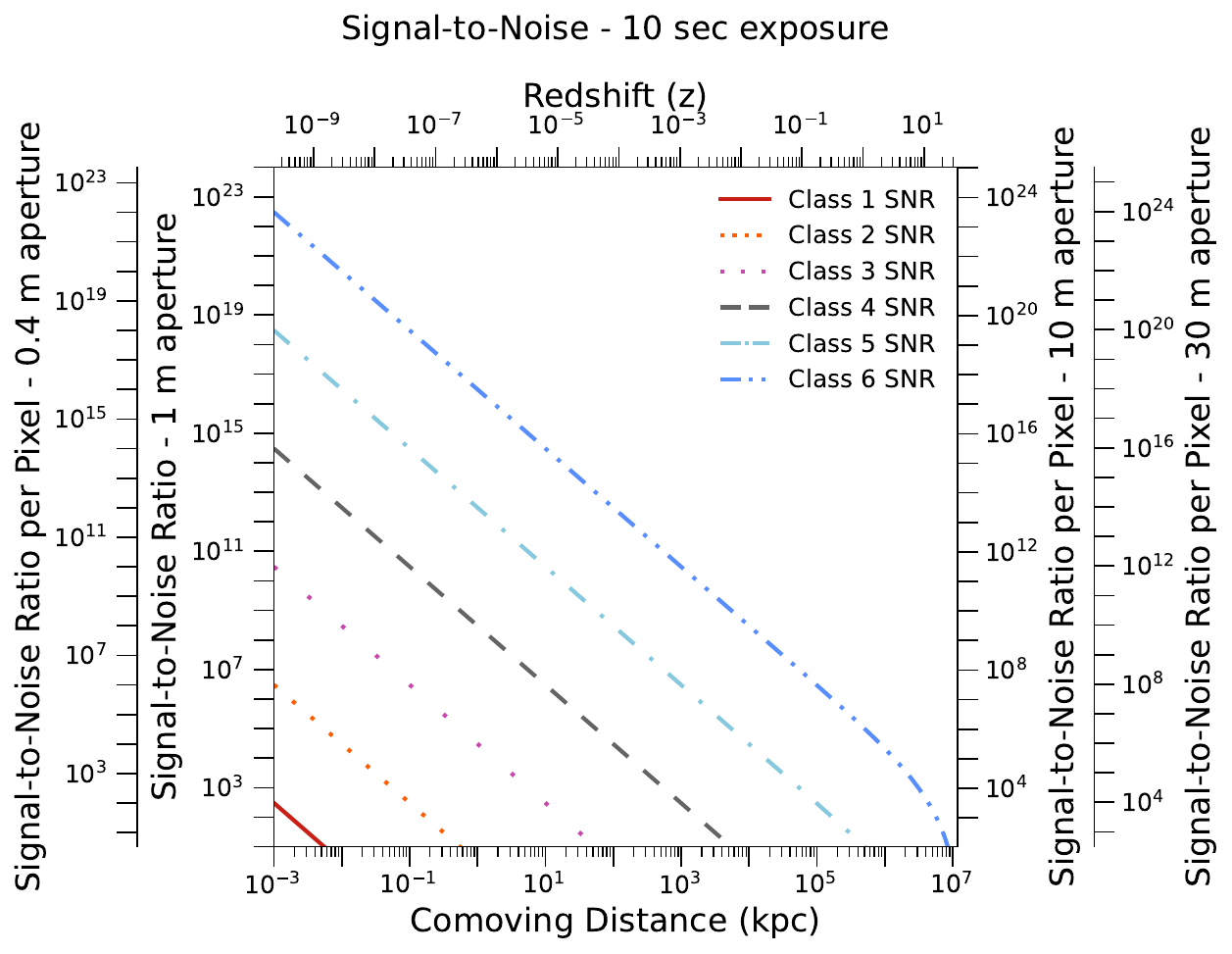}
    \caption{Signal-to-noise vs. distance while varying civilization class and aperture size. At high redshifts, flux scales as $(1+z)^{-2}$, due to the combined effects of redshift and the reduced photon arrival rate. We utilize LCOGT 0.4 m parameters as listed in Table \ref{table:LGTS} with various aperture sizes. Using the scales, we can determine the minimum civilization class that can be detected for a given distance and aperture size. If only a single pulse is received following the intelligent targeting assumption, the integration time increases the time-dependent noise but not the signal, and the signal-to-noise decreases by a factor of the pulse duration over the exposure time (e.g., $10^4$ for a millisecond). Redshift is calculated with Lambda cold dark matter ($\Lambda$CDM) cosmological parameters $H_0=\SI{67.4}{\km\per\s\per\mega\parsec}$, $\Omega_m=0.315$, $\Omega_\Lambda=0.685$ (\cite{Planck2020}). Figure adapted with permission from \cite{Lubin2016}.}
    \label{SNR}
\end{figure}

\begin{figure}[hbt]
    \centering
    \includegraphics[width = \columnwidth]{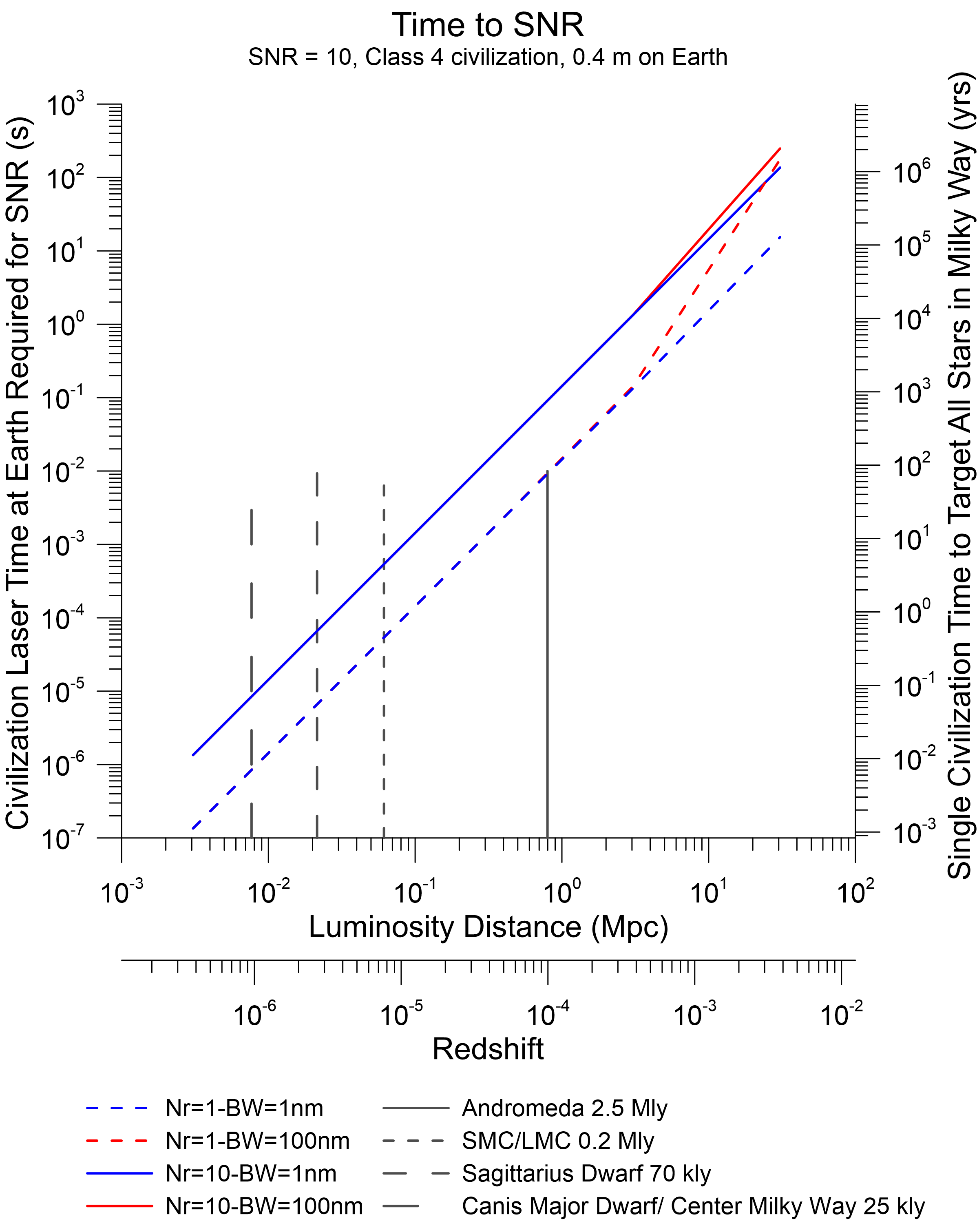}
    \caption{Time to SNR vs Luminosity Distance with low background magnitude (${\sim}21$ mag/arcsec$^2$) in the background noise dominated regime with LCOGT 0.4 m telescope parameters. This background corresponds to about 10 kpc from the nucleus for the \texttt{R} band as shown in Figure \ref{brightness}. For a civilization class 4 laser in a nearby galaxy, an LCOGT 0.4 m telescope network reaches an SNR ratio of 10 within about 10 ms of exposure time.}
    \label{Time to SNR}
\end{figure}

Figure \ref{SNR} summarizes the results of SNR as a function of several apertures resulting in high signal-to-noise even at large distances and long exposure times. The noise has two components: readout noise and time-dependent noise. Readout noise dominates for short integration times while the time-dependent part dominates at longer integration times. This is expanded in Figure \ref{Time to SNR} which plots a 10 second blind integration and shows it is not unreasonable for a SETI source to be detected even at vast distances using modest searches. 

{}

The integration time $\tau$ for a given SNR is 
\begin{align}
\tau = \frac{S_N^2n_t^2}{2F^2A_\epsilon^2} 
\left(1 + \sqrt{1 + \frac{4F^2A_\epsilon^2N_R^2}{S_N^2N_T^2}} \right)
\end{align}
where the time-dependent noise $n_t(e^-s^{-1/2})$ is
\begin{equation}
    n_t^2 = i_{DC} + F_BA_\epsilon\Omega
\end{equation}

\noindent and the total noise $N_T$ is

\begin{equation}
    N_T^2 = N_R^2 + \tau n_t^2.
\end{equation}

As one noise mode will dominant, it is often helpful to think of the noise as containing two regimes with the transition time $\tau_c$ occurring at equality: 
\begin{align}
\tau_c = \frac{N_R^2}{n_t^2} = \frac{N_ R^2}{FA_\epsilon + i_{DC} + F_BA_\epsilon \Omega} \approx \frac{N_R^2}{i_{DC} + F_BA_\epsilon \Omega}.
\end{align}

{}

\begin{figure}[hbt!]
    \centering
    \includegraphics[width = \columnwidth]{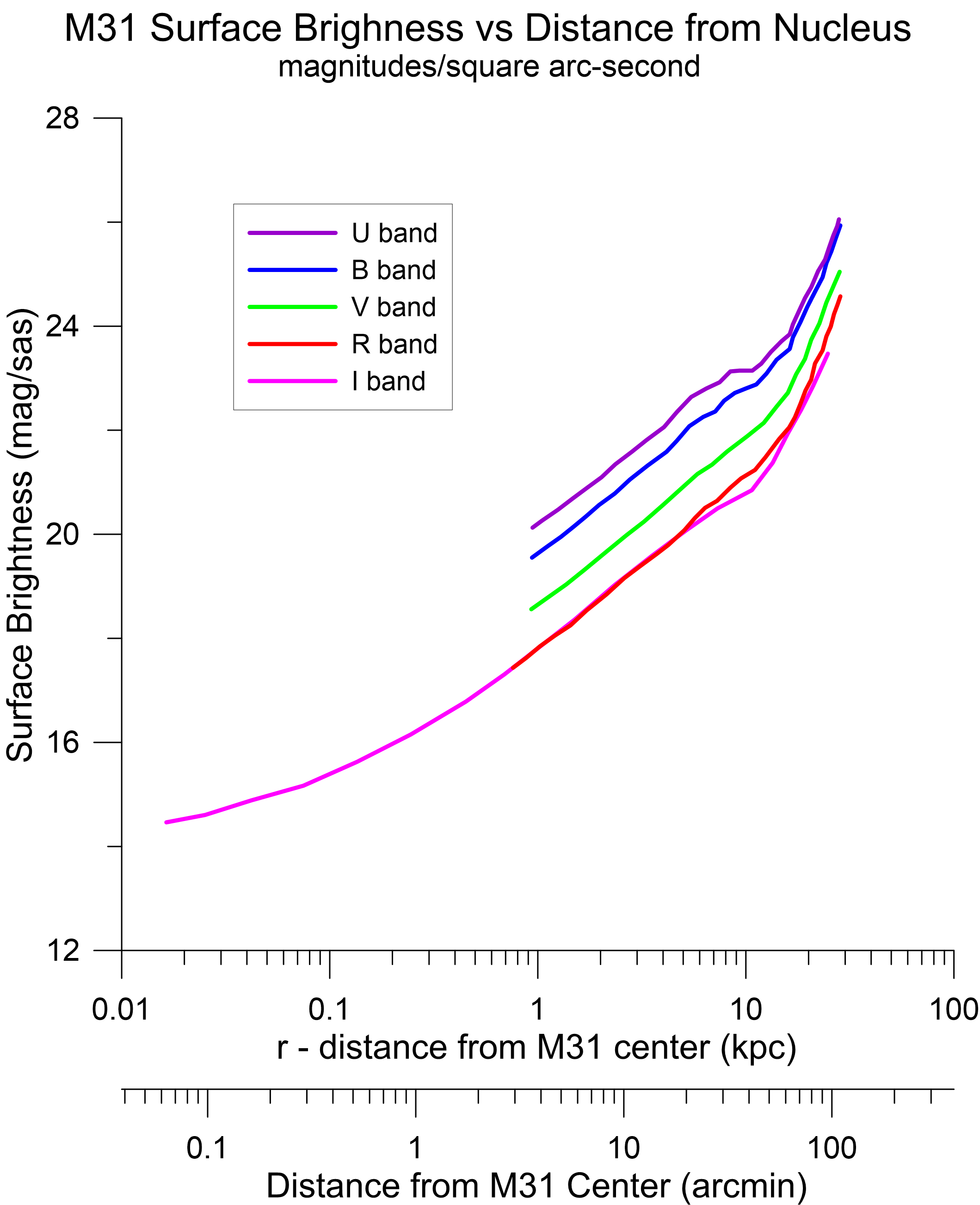}
    \caption{Surface brightness of M31 as a function of the radial distance from the nucleus for various band filters. Adapted from \cite{Courteau2011}.}
    \label{brightness} 
\end{figure}{}

Following the intelligent targeting assumption, we calculate the required time to target all stars in the Milky Way. Ignoring pointing time, a laser array in M31 could target all stars in the Milky Way in ${\sim}100$ years. A class 4 civilization in SMC or LMC could target all Milky Way stars in just a few years. If we choose to conduct a serious METI effort, the time to target all stars in a galaxy would be similar. We also emphasize that a laser array in the Milky Way center could target all Milky Way stars in months under these assumptions.

\section{Local Galactic Transient Survey} \label{LGTS}
\noindent As a successor to our Trillion Planet Survey (TPS; \cite{Stewart2017}), the Local Galactic Transient survey aims to survey M31, the Large Magellanic Cloud (LMC), and the Small Magellanic Cloud (SMC) for laser beacons from an intelligent civilization via LCOGT network's telescope array of 0.4 m telescopes. The Magellanic Clouds fit similar imaging criteria to M31 because they are nearby, dense regions. For this expansion, we created the LGTS from our previous trillion planet survey which focused solely on M31. Observations primarily focused on M31 due to its large size, favorable relative angle, density of stars, and proximity.

An integration time of 10 seconds is used for the majority of data. As shown in Figure \ref{Time to SNR}, 10 seconds significantly exceeds the required time for a LCOGT 0.4 m telescope to detect a civilization class $S\geq4$ laser. This short exposure time minimizes the background noise and required observing time while still providing a reasonable SNR ratio for stellar transient detection. 

Based on the calculations in Section \ref{SETI}, the expected magnitude of a directed energy source from a class 4 civilization (defined by \cite{Lubin2016}) would be readily detectable by LCOGT 0.4 m telescopes as seen in Figures \ref{SNR} and \ref{Time to SNR}. 

Each observation request takes 100 images with an LCOGT 0.4 m telescope. Science images are taken without filters so detection of transients across the optical and near-infrared spectrum is possible. Periodic transient sources can then be observed with filters to determine more about their characteristics. \\

\begin{deluxetable}{ll}[htb]
\tablecolumns{2}
\tablecaption{LGTS and LCOGT 0.4m Overview \label{table:LGTS}}
\tablehead{\colhead{Parameter} & \colhead{Value}}
\startdata
    \textit{LGTS Survey Overview}                                           \\
    Estimated Seeing ($^{\prime\prime}$)& 1                                 \\ \hline
    Integration time $\tau$ (s)         & 10                                \\ \hline
    Images per observation              & 100                               \\ \hline
    Sections Imaged                     & 78                                \\ \hline
    Total Science Images                & 29,753                            \\ \hline\hline
    \textit{Instrumental Specifications}                                    \\
    Primary mirror diameter (m)         & 0.4                               \\ \hline
    FOV                                 & 29.2$^{\prime}$ x 19.5$^{\prime}$ \\ \hline
    Pixel size ($^{\prime\prime}$)      & 0.571 arcsec                      \\ \hline
    Image size (pixels)                 & 3,000 x 2,000                     \\ \hline
    Readout noise $N_R$ ($\unit{e^-}$)  & 14.5                              \\ \hline
    Gain ($\unit{e^- ADU^{-1}}$)        & 1.6                               \\ \hline
    Dark current $i_{DC}$ ($\unit{e^-pixel^{-1} s^{-1}}$) & 0.03            \\ \hline
    Quantum efficiency $Q_e$            & 50\%\text{ at 500 nm}             \\ \hline
    Assumed optical efficiency $\epsilon$ (\%)            & 50              \\ \hline
\enddata
\end{deluxetable}
\vspace{-1.3cm}

\begin{figure}[h!]
    \centering
    \includegraphics[width = \columnwidth]{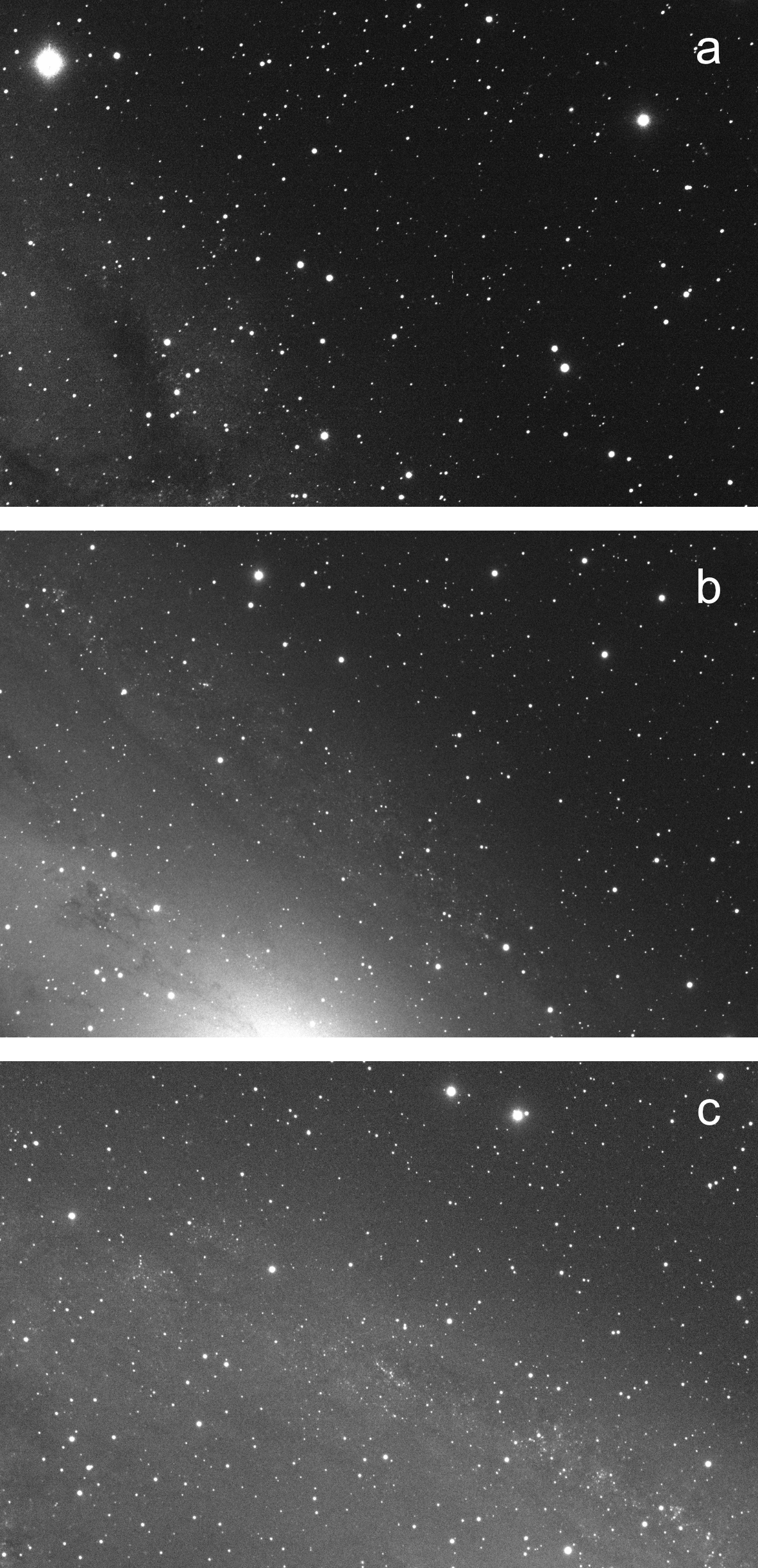}
    \caption{The wide range of surface brightness in LGTS data with untypically long exposures for visibility. a) LGTS Section 38 taken at LCOGT's McDonald Observatory on 2019 October 18 with an integration time of 60 s. b) LGTS Section 23 taken at LCOGT's McDonald Observatory on 2019 July 19 with an exposure time of 100 s. c) LGTS Section 24 of M31 taken at LCOGT's Haleakala Observatory on 2019 August 18 with an integration time of 100 s.}
    %\vspace{0.5in}
    \label{images}
\end{figure}

\begin{figure*}[t]
    \centering
    \includegraphics[height = 2.2in]{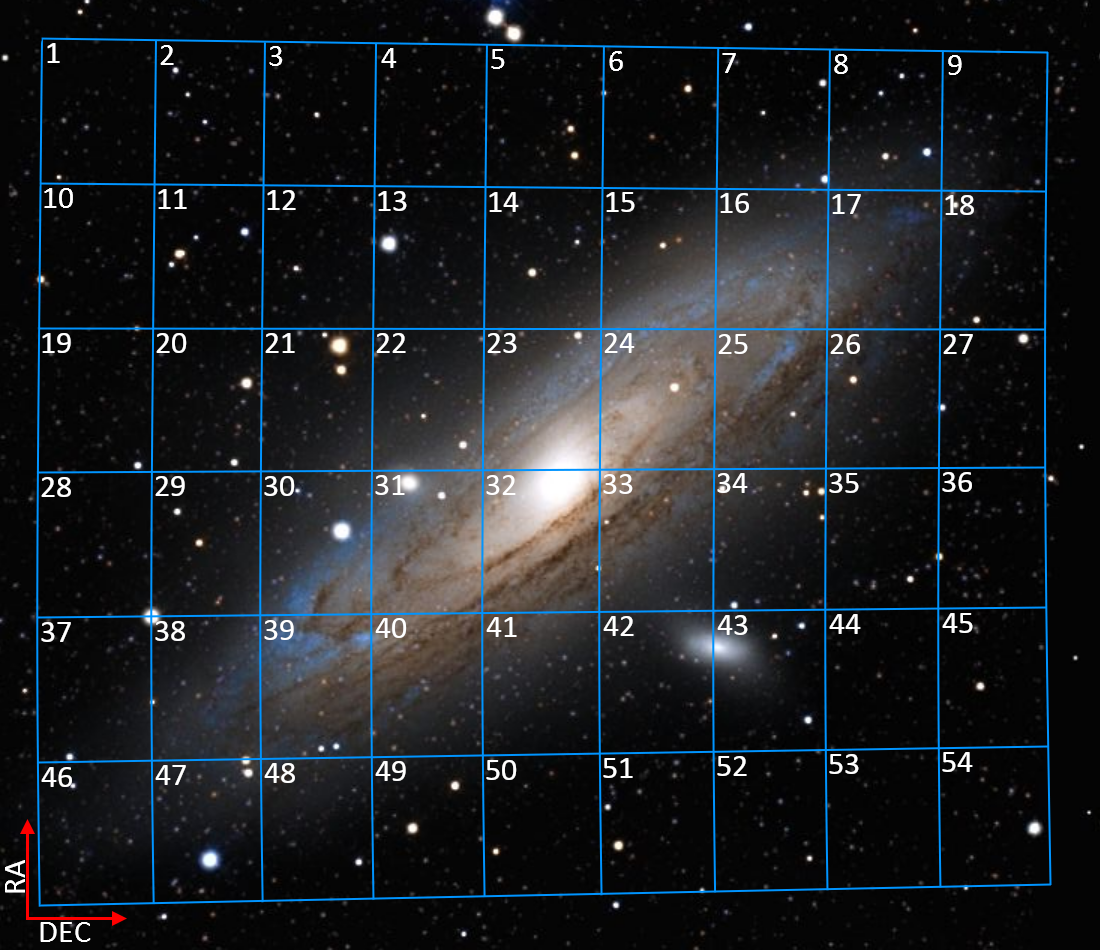} \hfill
    \includegraphics[height = 2.2in] {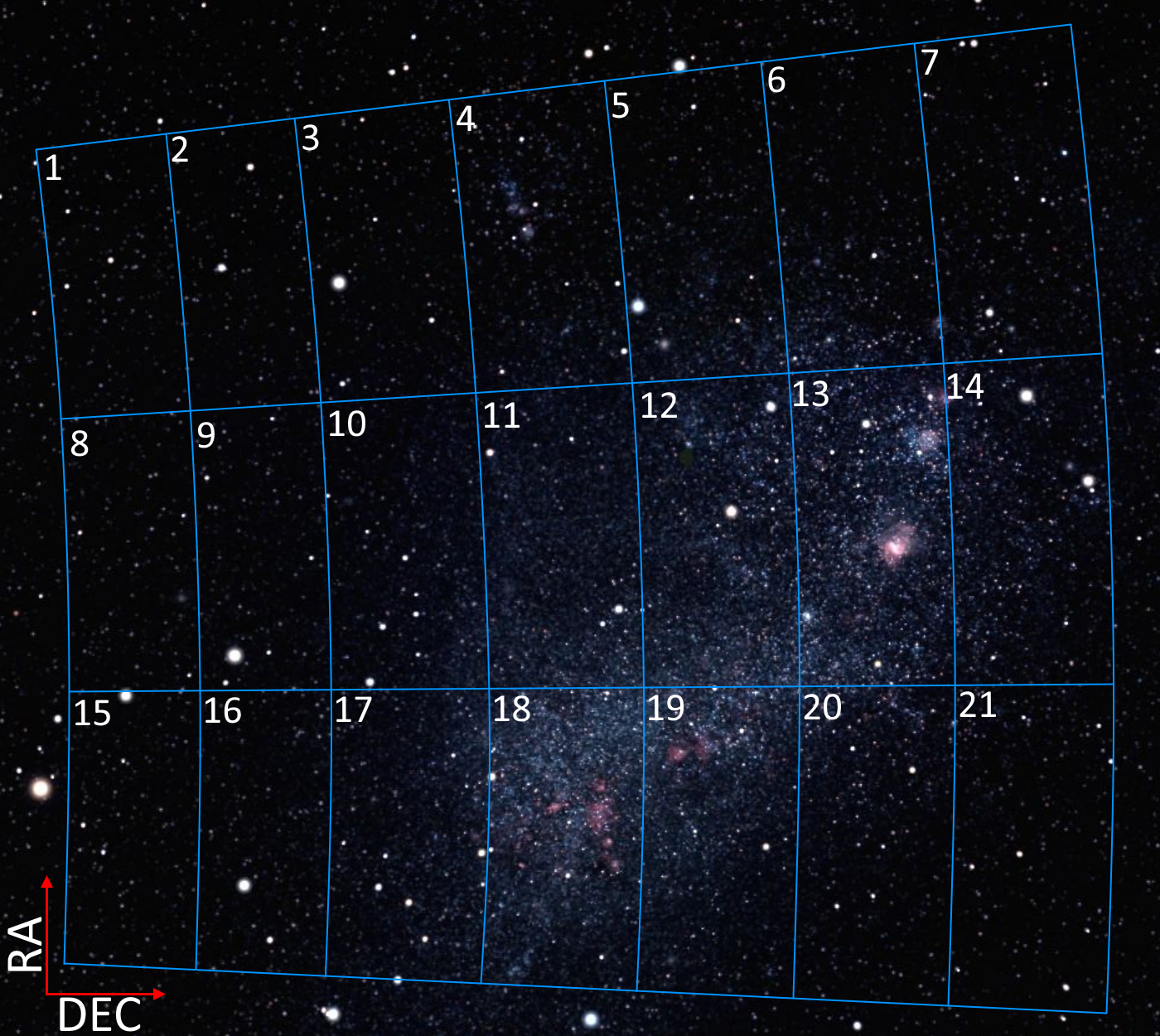} \hfill
    \includegraphics[height = 2.2in]{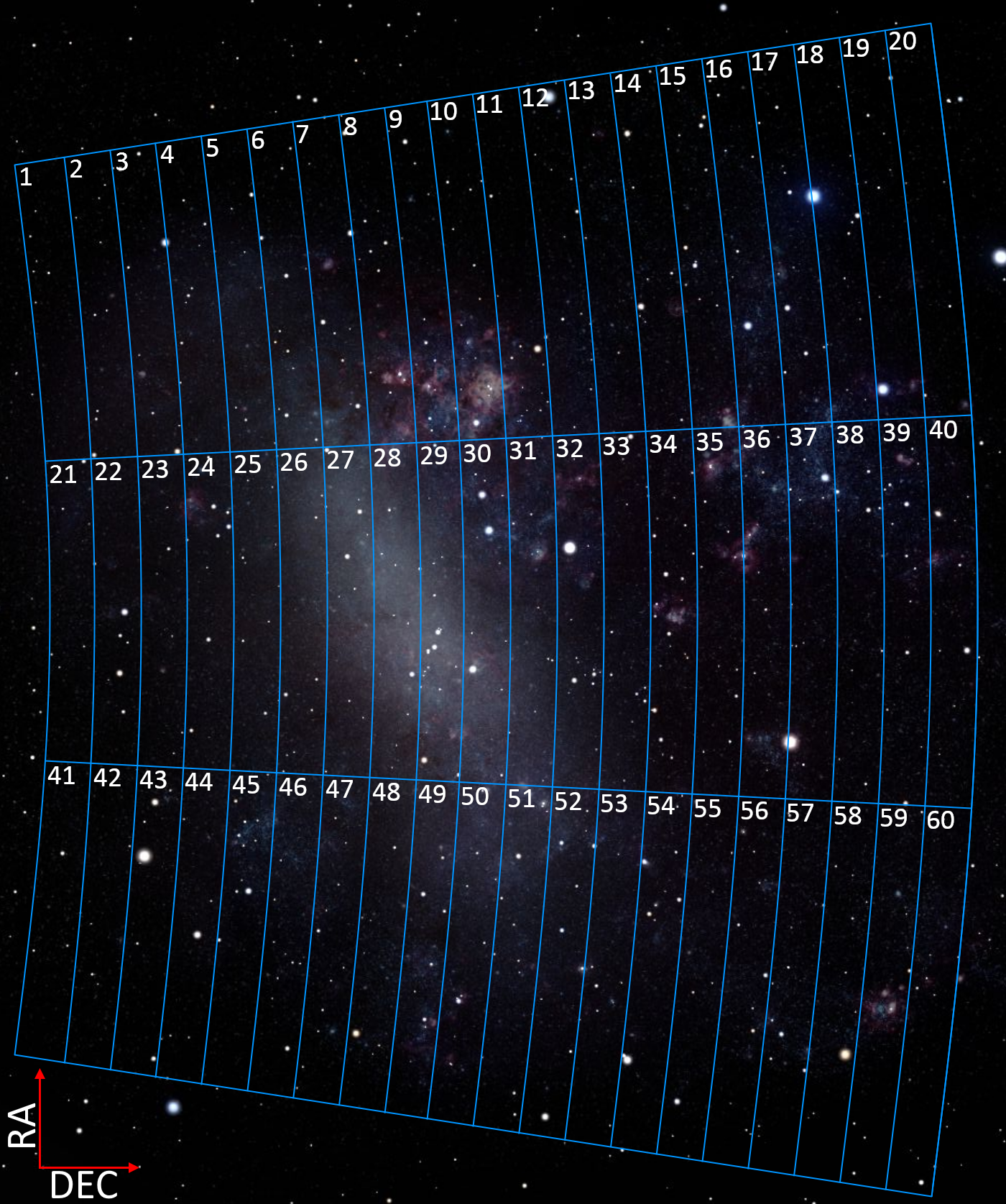}
    \caption{
    M31, SMC, and LMC survey sections. Only sections with galactic structure were imaged by the LGTS. Galactic images generated in Stellarium using a small FOV Mercator projection.}
    \label{sections}
\end{figure*}

As LCOGT 0.4 m telescopes have a smaller field of view (29.2$'$ x 19.5$'$) than the local galaxies that cover patches of sky in the degree range, we sectioned M31, SMC, and LMC as shown in Figures \ref{images} and \ref{sections}. We do not image sections that do not contain the target galaxy. The 2023 upgrade of LCOGT's 0.4 m network enables wide-field imaging with a vastly improved 1.9$'$ x 1.3$'$ field of view (15x). This upgrade will greatly decrease the required number of sections required to survey the local galaxies and expedite the rate of future surveys \cite{Harbeck2023}.

{}

{}

Data for this project is collected in batches, imaging 3-5 sectors approximately every two weeks to build a catalog with regular differences in its time domain. This schedule allows us to construct and compare light curves from data taken over different parts of their phase. We have collected 29,753 images from 78 unique LGTS sections with a storage requirement of 518 GB.

Preprocessing occurs in LCOGT network's BANZAI pipeline which applies overscan, gain, and bias subtractions along with flat-field correction, source detection, and astrometry. The full details are available in LCOGT network's documentation on BANZAI (\cite{banzai}). Preprocessed data can be seen in Figure \ref{images} showing the varied background magnitude as a function of radial distance from the galactic center.

Data processing is done by the near-real time TRansient Image Processing Pipeline (TRIPP; \cite{TRIPP}). The TRIPP pipeline is an open-source transient and variable source detection pipeline with difference imaging analysis and light curve analysis techniques. TRIPP was programmed specifically with LGTS in mind, however, the pipeline and its optimizations could be broadly beneficial for Time-domain Astrophysics. These techniques are used in recent optical surveys such as \cite{Richmond2019}, \cite{Bonanos2019}, \cite{Moretti2018}, \cite{Morganson2018}, and \cite{Jencson2019}. Following these methods, TRIPP could feasibly detect OSETI signals in LGTS data.

\section{Conclusion} \label{conclusion}
\noindent Our calculations indicate that with modest surveys, a civilization within M31 utilizing laser technology that we could construct in this century would be readily detectable with the LGTS's observational capabilities and TRIPP's detection abilities. Thus, a SETI detection from a directed energy source following intelligent targeting would be best enabled by high cadence large sky surveys carried out by a large collaboration or many groups in parallel. Early processing has validated TRIPP's performance is consistent with other photometric processing pipelines with 10 successful observing nights of SN2023ixf spanning  May 25th through July 11th, 2023 (\cite{TRIPP}).

LGTS data collection, TRIPP pipeline development, and TRIPP pipeline validation using LGTS data have been completed. We are now reprocessing all of the nearly 30,000 LGTS images collected during our 5-year survey phase with the finalized pipeline. Image processing of LGTS images via TRIPP occurs at ${\sim}$6 seconds per image with the photometry of transient candidates to be reported in a forthcoming paper.

\section*{Acknowledgements}
\noindent The repository used to generate a majority of our figures is available on Zenodo (\cite{lgts-plots}). We thank the many undergraduates who worked on this project before us. In addition, this work uses observations from the Las Cumbres Observatory Global Telescope Network, and we are grateful for the support of the network and their generous allotments of observing time. 
Funding for this project comes from the UCSB Faculty Research Assistance Program and Undergraduate Research and Creative Activities grants, NASA grants: NIAC Phase I DEEP-IN – 2015 NNX15AL91G and NASA NIAC Phase II DEIS – 2016 NNX16AL32G, the NASA California Space Grant (NASA NNX10AT93H), as well as a generous gift from the Emmett and Gladys W. fund.

\bibliographystyle{AASJournal}
\bibliography{Bib.bib} 
\end{document}